\documentclass[12pt,letterpaper]{article}

\usepackage[utf8]{inputenc}
\usepackage[english]{babel}
\usepackage{amsmath}
\usepackage{amsfonts}
\usepackage{amssymb}
\usepackage{graphicx}
\usepackage[left=1in,right=1in,top=1in,bottom=1in]{geometry} 
\usepackage{changepage} 
\usepackage{times} 
\usepackage{titlesec} 
\usepackage{cite} 
\usepackage[colorlinks=true,citecolor=cyan,linkcolor=red,bookmarks=true,bookmarksnumbered=true]{hyperref} 
\usepackage{tikz} 
\usetikzlibrary{arrows.meta} 
\usepackage[font=footnotesize,labelfont=bf]{caption} 
\usepackage{soul}
\usepackage[mathscr]{eucal}

\renewcommand{\thesection}{\Roman{section}}
\renewcommand{\thesubsection}{\arabic{subsection}}

\titleformat{\section}[hang]{\bf\large}{\thesection\enspace}{5pt}{}
\titleformat{\subsection}[hang]{\bf\normalsize}{\thesubsection\enspace}{5pt}{}

\begin{document}

{
\flushleft\Huge\bf Graphene in complex magnetic fields
}\vspace{5mm}

\begin{adjustwidth}{1in}{} 
{
\flushleft\large\bf David J Fern{\'a}ndez C and Juan D Garc{\'i}a-Mu{\~n}oz\footnote{Author to whom correspondence should be addressed}
}
\vspace{2.5mm}
{
\par\noindent\small Physics Department, Cinvestav, P.O.B. 14-740, 07000 Mexico City, Mexico
}
\vspace{2.5mm}
{
\par\noindent\small Email: david@fis.cinvestav.mx and dgarcia@fis.cinvestav.mx  
}
\vspace{5mm}
{
\par\noindent\bf Abstract
}
{ 
\newline\small Exact analytic solutions for an electron in graphene interacting with external complex magnetic fields are found. The eigenvalue problem for the non-hermitian Dirac-Weyl Hamiltonian leads to a pair of intertwined Schr{\"o}dinger equations, which are solved by means of supersymmetric quantum mechanics. Making an analogy with the non-uniform strained graphene a prospective physical interpretation for the complex magnetic field is given. The probability and currents densities are explored and some remarkable differences as compared with the real case are observed.  
}
\vspace{2mm}
\end{adjustwidth}

\section{Introduction}

At the beginning of the twenty-first century a $2D$-material known as graphene was isolated for the first time by Geim and Novoselov \cite{Novoselov2004}. After this discovery a lot of work trying to delve into its properties have been done. In particular, its electronic properties are of great interest, as the integer quantum Hall effect where the charge carriers behave as massless chiral quasiparticles with a linear dispersion relation. Within this wide field of study, the work of Kuru, Negro and Nieto is worth being mentioned \cite{Kuru2009}, since they show how to use supersymmetric quantum mechanics (SUSY QM) in order to find exact analytic solutions for a class of Hamiltonians, which describe the interaction of the electron in graphene with external magnetic fields. Other authors have explored further this technique, finding as well interesting results \cite{Milpas2011,Midya2014,Erik2017,Schulze2017,Concha2018,Roy2018,Erik2019,Celeita2020,Erik2020}.         
However, as far as we know a research about generalizing the method to complex magnetic fields has not been done yet.  

Even though complex magnetic fields are considered to be non-physical, in recent years it has been found that their effective action on the coherence of many-body quantum systems can be detected \cite{Peng2015}. Motivated by this result, in this paper we shall study from a theoretical viewpoint the effects caused on the electron behavior by a complex magnetic field applied orthogonally to the graphene surface, and how this generalization modifies the energies, probability and current densities as compared with the real case. We shall discuss as well a possible physical interpretation of such complex magnetic fields. 

The paper has been organized as follows: in section~\ref{S2} we will describe how SUSY QM works to solve the eigenvalue problem for the effective Hamiltonian of graphene in a complex magnetic field; section~\ref{S3} shows some cases where different complex magnetic profiles are taken, and for which the algorithm of the previous section can be applied. A discussion about a possible physical interpretation of complex magnetic fields, and an important effect induced by their non-null imaginary parts, is given in section~\ref{S4}; finally, in section \ref{S5} we present our conclusions.

\section{Effective Hamiltonian for graphene in complex magnetic fields} \label{S2}

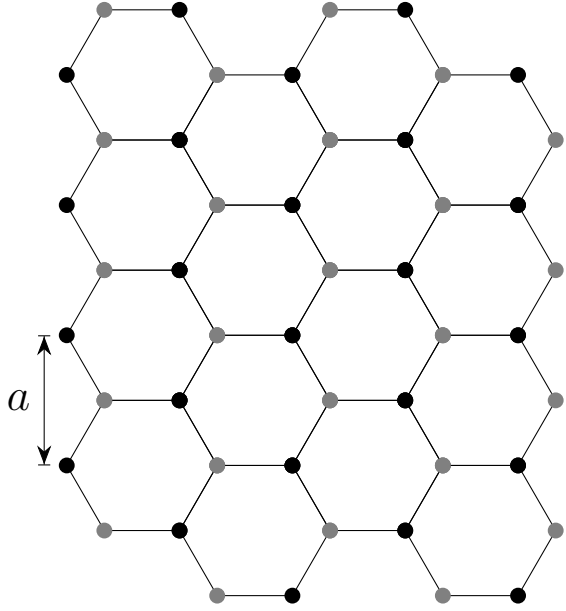
\begin{figure}[t]
\begin{center}
\begin{tikzpicture}

\foreach \n in {0,1,2,3}
{
\ifodd \n \newcommand{\xo}{3/2*\n} \newcommand{\yo}{-0.866}
\else \newcommand{\xo}{3/2*\n} \newcommand{\yo}{0} 
\fi    
\foreach \c in {0,1,2,3}
{
\foreach \x in {0,1,3,4} 
{
\foreach \y in {0,1,-1} 
{
\ifnum \x=0 \ifnum \y=0 \fill[color=black] (\xo+\x/2,\yo+\y*0.866+\c*1.732) circle [radius=3pt]; 
\draw (\xo+\x/2+1/2,\yo+\y*0.866-0.866+\c*1.732)--(\xo+\x/2,\yo+\y*0.866+\c*1.732)--(\xo+\x/2+1/2,\yo+\y*0.866+0.866+\c*1.732); 
\fi \fi
 \ifnum \x=1 \ifnum \y=1 \draw (\xo+\x/2,\yo+\y*0.866+\c*1.732)--(\xo+\x/2+1,\yo+\y*0.866+\c*1.732); \fill[color=gray] (\xo+\x/2,\yo+\y*0.866+\c*1.732) circle [radius=3pt]; 
 \fi
 \ifnum \y=-1 \draw (\xo+\x/2,\yo+\y*0.866+\c*1.732)--(\xo+\x/2+1,\yo+\y*0.866+\c*1.732); \fill[color=gray] (\xo+\x/2,\yo+\y*0.866+\c*1.732) circle [radius=3pt]; 
 \fi \fi
 \ifnum \x=3 \ifnum \y=1 \draw (\xo+\x/2,\yo+\y*0.866+\c*1.732)--(\xo+\x/2+1/2,\yo+\y*0.866-0.866+\c*1.732); \fill[color=black] (\xo+\x/2,\yo+\y*0.866+\c*1.732) circle [radius=3pt]; \fi
 \ifnum \y=-1 \draw (\xo+\x/2,\yo+\y*0.866+\c*1.732)--(\xo+\x/2+1/2,\yo+\y*0.866+0.866+\c*1.732); \fill[color=black] (\xo+\x/2,\yo+\y*0.866+\c*1.732) circle [radius=3pt]; \fi \fi
 \ifnum \x=4 \ifnum \y=0 \fill[color=gray] (\xo+\x/2,\yo+\y*0.866+\c*1.732) circle [radius=3pt]; 
 \fi \fi
}
}
}
}
\draw[arrows={|Stealth[scale=1.5]-Stealth[scale=1.5]|}] (-0.3,0)-- node[left=1pt]{\Large $a$} (-0.3,1.732); 
\end{tikzpicture}
\caption{Graphene structure. The intramolecular distance $a$ is traced. }\label{graphene}
\end{center}
\end{figure}

A hexagonal structure of carbon atoms in a honeycomb $2D$-arrangement is called monolayer graphene, or simply graphene (see figure \ref{graphene}). In the study of this material one usually works with an effective Hamiltonian describing the hopping of an electron from one atom to any of its nearest neighbors \cite{Katnelson2011,Mermin1976,Raza2012,Saito1998}. Such Hamiltonian can be written as a $2\times 2$ matrix operator of the form
\begin{equation} \label{E-1}
H = v_{0}
\begin{pmatrix}
0 && \pi \\
\pi^{\dagger} && 0
\end{pmatrix},
\end{equation}  
where $v_{0} = \sqrt{3}a\gamma_{0}/2\hbar$ is the Fermi velocity and $\pi = p_{x}-ip_{y}$. The quantity $a \approx 2.46\ \textup{\r{A}}$ is the intramolecular distance in the graphene layer \cite{McCann2013}, while $\gamma_{0} \approx 3.033\ \text{eV}$ is known as the in-plane hopping parameter, which is equal to the negative binding energy between two adjacent carbon atoms \cite{Saito1998}. On the other hand, $p_{j}$ is the momentum operator in the $j$-th direction, with $j=x,y$. 

Let us suppose that a complex magnetic field orthogonal to the graphene layer is applied, which varies only along one direction, e.g., $\mathbf{B}(x) = B(x)\mathbf{e}_{z}$, $B(x)\in \mathbb{C}$. In the Landau gauge the associated vector potential can be written as $\mathbf{A}(x) = A(x)\mathbf{e}_{y}$, where $B(x) = dA(x)/dx$. Taking into account the minimal coupling rule, the effective Hamiltonian \eqref{E-1} becomes now
\begin{equation} \label{E-2}
H = v_{0}
\begin{pmatrix}
0 && p_{x} - ip_{y} - i\frac{e}{c}A(x) \\
p_{x} + ip_{y} + i\frac{e}{c}A(x) && 0
\end{pmatrix}.
\end{equation}
Since $H$ is invariant under translations along $y$-direction, its eigenvectors can be expressed as: 
\begin{equation} \label{E-3}
\Psi(x,y) = Ne^{iky}
\begin{pmatrix}
\psi^{+}(x) \\
i\psi^{-}(x)
\end{pmatrix},
\end{equation} 
with $N$ being a normalization factor and $k$ the wavenumber in $y$-direction. In the coordinates representation the momen{\-t}um operator $p_{j}$ can be written as $-i\hbar \partial_{j}$, with $\partial_{j} = \partial/\partial{j}$, $j = x,y$, thus the eigenvalue equation for $H$ looks like
\begin{equation} \label{E-4}
H\Psi(x,y) = \hbar v_{0}
\begin{pmatrix}
0 && -i\partial_{x} - \partial_{y} - i\frac{e}{c\hbar}A(x) \\
-i\partial_{x} + \partial_{y} + i\frac{e}{c\hbar}A(x) && 0
\end{pmatrix} \Psi(x,y) = E\Psi(x,y). 
\end{equation}
Using expression~(\ref{E-3}), the matrix equation \eqref{E-4} is reduced to a coupled system of equations:
\begin{equation} \label{E-5}
\begin{aligned}
& L^{-}\psi^{-}(x) \equiv \left[\frac{d}{dx} + k + \frac{e}{c\hbar}A(x)\right]\psi^{-}(x) = \mathcal{E}\psi^{+}(x), 
\\
& L^{+}\psi^{+}(x) \equiv  \left[-\frac{d}{dx} + k + \frac{e}{c\hbar}A(x)\right]\psi^{+}(x) = \mathcal{E}\psi^{-}(x), 
\end{aligned}
\end{equation}
with $\mathcal{E} = E/\hbar v_{0}$. It is important to realize that $L^{+}$ is not the Hermitian conjugate of $L^{-}$, $\left(L^{-}\right)^{\dagger} = -d/dx + k + \frac{e}{c\hbar}\bar{A}(x)\neq L^{+}$, with $\bar{z}$ being the complex conjugate of $z$. In order to decouple the system~\eqref{E-5}, let us apply $L^{+}$ on the first equation and $L^{-}$ on the second, which leads to:
\begin{equation} \label{E-6}
\begin{aligned}
& L^{+}L^{-}\psi^{-}(x) = \left[-\frac{d^{2}}{dx^{2}} + \left(k + \frac{e}{c\hbar}A(x)\right)^{2} - \frac{e}{c\hbar}A'(x)\right]\psi^{-}(x) = \varepsilon \psi^{-}(x),
\\
& L^{-}L^{+}\psi^{+}(x) = \left[-\frac{d^{2}}{dx^{2}} + \left(k + \frac{e}{c\hbar}A(x)\right)^{2} + \frac{e}{c\hbar}A'(x)\right]\psi^{+}(x) = \varepsilon\psi^{+}(x),
\end{aligned}
\end{equation} 
where $\varepsilon=\mathcal{E}^{2}$ and $dA(x)/dx \equiv A'(x)$. It is natural to identify now
\begin{equation}\label{E-6a}
H^- = L^{+}L^{-}, \qquad H^+ = L^{-}L^{+},
\end{equation}
with $H^\pm$ being two non-Hermitian Hamiltonians fulfilling
\begin{equation} \label{E-7}
H^+L^{-} = L^{-}H^{-}.
\end{equation}
The intertwining relation~\eqref{E-7}, together with the expressions \eqref{E-5} for $L^{\pm}$ and the factorizations in equation~\eqref{E-6} are the basis of the so-called supersymmetric quantum mechanics (SUSY QM).
In fact, it is standard to denote 
\begin{equation} \label{E-9}
L^{\pm} =\mp \frac{d}{dx} + \text{w}(x),
\end{equation}       
with the complex function 
\begin{equation} \label{superpotential}
\text{w}(x) = k + \frac{e}{c\hbar}A(x)
\end{equation}
being called superpotential. Thus, the non-Hermitian Hamiltonians $H^\pm$ take the form
\begin{equation} \label{E-8}
H^{\pm} = -\frac{d^{2}}{dx^{2}} + V^{\pm}(x),
\end{equation}
where the complex SUSY partner potentials $V^\pm$ are written in terms of the superpotential as follows:
\begin{equation} \label{E-10}
V^{\pm} = \text{w}^{2}(x) \pm \text{w}'(x).
\end{equation}   

Suppose now that $\psi^{\pm}_{n}(x)$ are eigenfunctions of $H^{\pm}$ with eigenvalues $\varepsilon^{\pm}_{n}$, the quantum number $n$ being a non-negative integer. We choose $H^-$ as the Hamiltonian having the null energy as one of its eigenvalues, i.e., $\varepsilon^{-}_{0}=0$. This automatically fixes the superpotential, since
\begin{equation}
H^- \psi^{-}_{0} = \varepsilon^{-}_{0} \psi^{-}_{0} = 0 \quad \Rightarrow \quad L^- \psi^{-}_{0} = 0 = (\psi^{-}_{0})' + \text{w}(x) \psi^{-}_{0} \quad \Rightarrow \quad \text{w}(x)= - \frac{(\psi^{-}_{0})'}{\psi^{-}_{0}},
\end{equation}
where equation~\eqref{E-6a} was used. As $\psi^{-}_{0}$ is square-integrable, the solution to $H^+ \psi^+=0$, which also satisfies $L^+ \psi^+ = 0$ $\Rightarrow$ $\psi^+=1/\psi^{-}_{0}$, is not square-integrable, thus $\varepsilon^{-}_{0}=0$ is not in the spectrum of $H^+$. However, the intertwining relation~\eqref{E-7} guarantees that any other non-null eigenvalue of $H^{-}$ ($\varepsilon^{-}_{n}, \, n=1,2,\dots$) belongs to the spectrum of $H^+$. Proceeding by analogy with the real case, we will denote $\varepsilon^{+}_{n-1}= \varepsilon^{-}_{n}$, thus the corresponding eigenstates $\psi^{\pm}_{n}(x)$ are interrelated through $L^\pm$ as follows: 
\begin{equation} \label{E-12}
\psi^{+}_{n-1}(x) = \frac{L^{-}\psi^{-}_{n}(x)}{\sqrt{\varepsilon^{-}_{n}}},\quad \psi^{-}_{n}(x) = \frac{L^{+}\psi^{+}_{n-1}(x)}{\sqrt{\varepsilon^{+}_{n-1}}}, \quad n=1,2,\dots
\end{equation}
Note that, despite $L^{+}$ is not the hermitian conjugate of $L^{-}$, the second equation in \eqref{E-12} is fulfilled since the factorizations \eqref{E-6a} imply that $H^{-}L^{+} = L^{+}H^{+}$, then $L ^{+}\psi^{+}_{n-1}(x)$ is an eigenfunction of $H^{-}$ with eigenvalue $\varepsilon^{-}_{n}$. 

It is important to stress that the potentials $V^\pm(x)$ are only auxiliary tools to solve the original problem, thus they do not have physical meaning. Moreover, they are typically shape-invariant SUSY partner potentials, since the factorization energy involved in equation~\eqref{E-6a} is the null energy associated to a Hamiltonian $H^{-}$ chosen as to have such a symmetry \cite{Gangopadhyaya2018}. 

Let us remark that the derivative of the superpotential is directly related to the magnetic field amplitude as follows
\begin{equation} \label{E-15}
\text{w}'(x) = \frac{e}{c\hbar}B(x).
\end{equation}

Coming back to our initial problem, the eigenvectors and eigenvalues of the Hamiltonian \eqref{E-2} describing the graphene layer in the complex magnetic fields are given by:
\begin{equation} \label{E-16}
\begin{aligned}
& \Psi_{0}(x,y) = e^{iky}
\begin{pmatrix}
0 \\
i\psi^{-}_{0}(x)
\end{pmatrix},\quad E_{0} = 0,
\\
& \Psi_{n}(x,y) = \frac{e^{iky}}{\sqrt{2}}
\begin{pmatrix}
\psi^{+}_{n-1}(x) \\
i\psi^{-}_{n}(x)
\end{pmatrix},\quad E_{n} = \pm\hbar v_{0}\sqrt{\varepsilon_{n}^{-}}, 
\end{aligned}
\end{equation}
with $n \in \mathbb{N}$. Let us mention that the energies $E_n$ with the plus sign are associated to electrons while the ones with the minus sign to holes. In our examples below only the electron energies will be considered. 

Before addressing some examples, let us define first two physical quantities that will help us to describe the electron behavior ruled by the Hamiltonian~\eqref{E-2}. Since such Hamiltonian is a piece of a $2\times 2$ diagonal-block supermatrix, whose non-zero extra block is the transpose of $H$, we are in fact dealing with a Dirac-like problem \cite{Katnelson2011}. However, it is sufficient to solve $H$ in order to obtain the whole solution of the Dirac-like Hamiltonian characterizing the monolayer graphene. The first physical quantity to be explored is the probability density defined by
\begin{equation} \label{E-17}
\rho = \Psi^{\dagger}\Psi,
\end{equation}  
while the second one is the probability current written as 
\begin{equation} \label{E-18}
\mathbf{J} = v_{0}\Psi^{\dagger}\vec{\sigma}\Psi.
\end{equation}
The previous expression for $\mathbf{J}$ is similar to the real case given in \cite{Kuru2009} and to the free case in \cite{Ferreira2011}, nevertheless the continuity equation turns out to be inhomogeneous, with the term of inhomogenei{\-t}y being given by
\begin{equation} \label{E-19}
\frac{2ev_{0}}{c\hbar}\text{Im}[A(x)]\Psi^{\dagger}\sigma_{y}\Psi.
\end{equation} 

\begin{figure}[t]
\begin{center}
\includegraphics[scale=0.5]{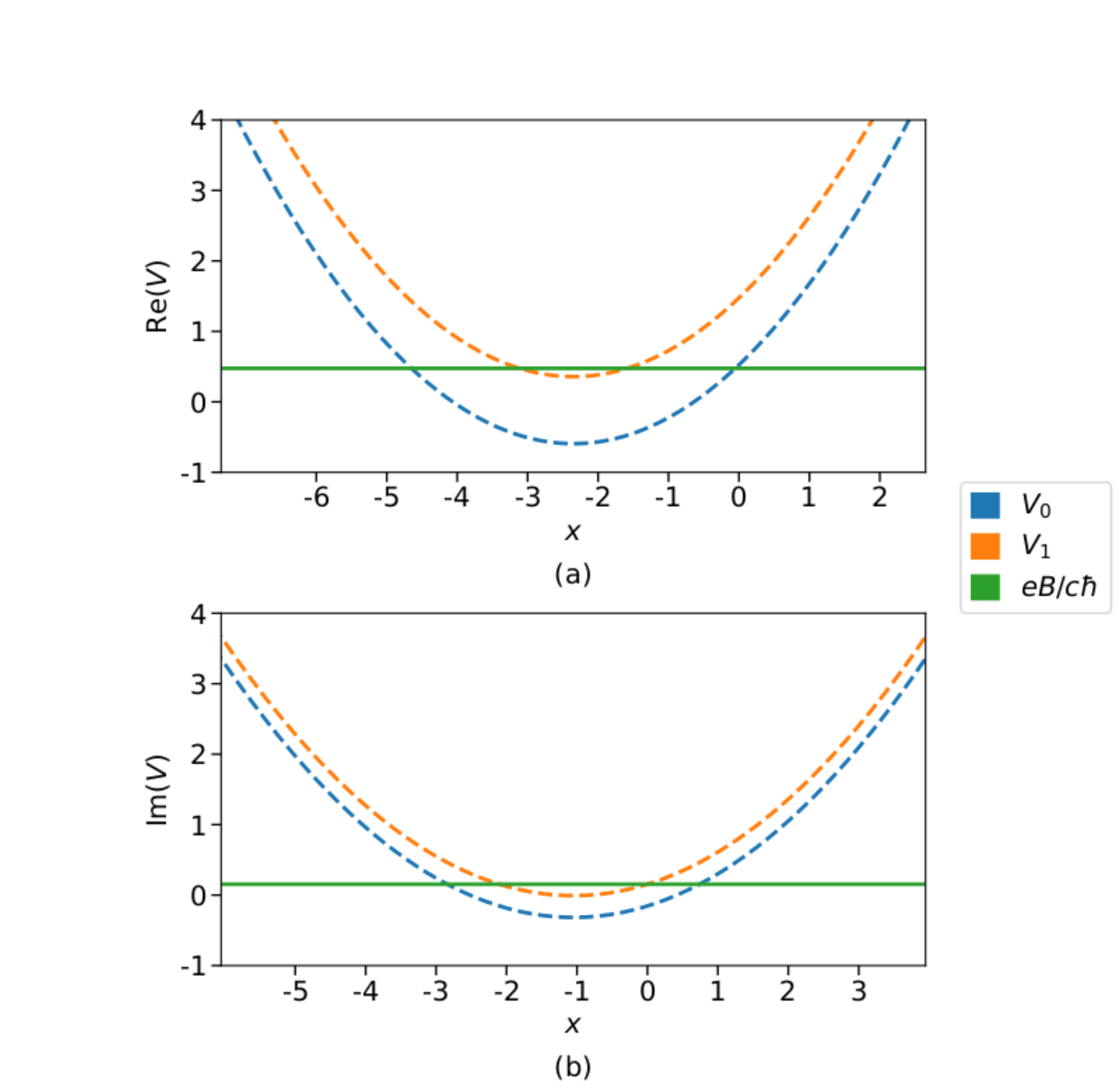}
\caption{Real (a) and imaginary part (b) of the complex harmonic oscillator potentials $V^\pm$ and the associated magnetic field. The parameters chosen are $|\omega|$ = $k$ = 1 and $\theta$ = $\pi/10$.} \label{F-1}
\end{center}
\end{figure}   

\section{Exactly solvable cases} \label{S3}

In these examples we shall take several magnetic profiles whose amplitude is the product of a complex constant times a real function of $x$. We shall determine the corresponding superpotential, the auxiliary SUSY partner potentials and then the solutions to the original problem. It is worth noting that we shall solve first the potential $V^{-}(x)$ and then, from its eigenfunctions and eigenva{\-l}ues, the corresponding solutions of $V^{+}(x)$ will be found. Moreover, any other parameter of the magnetic profile is supposed to be positive, unless otherwise specified. 

\subsection{Constant magnetic field}

\begin{figure}[t]
\begin{center}
\includegraphics[scale=0.5]{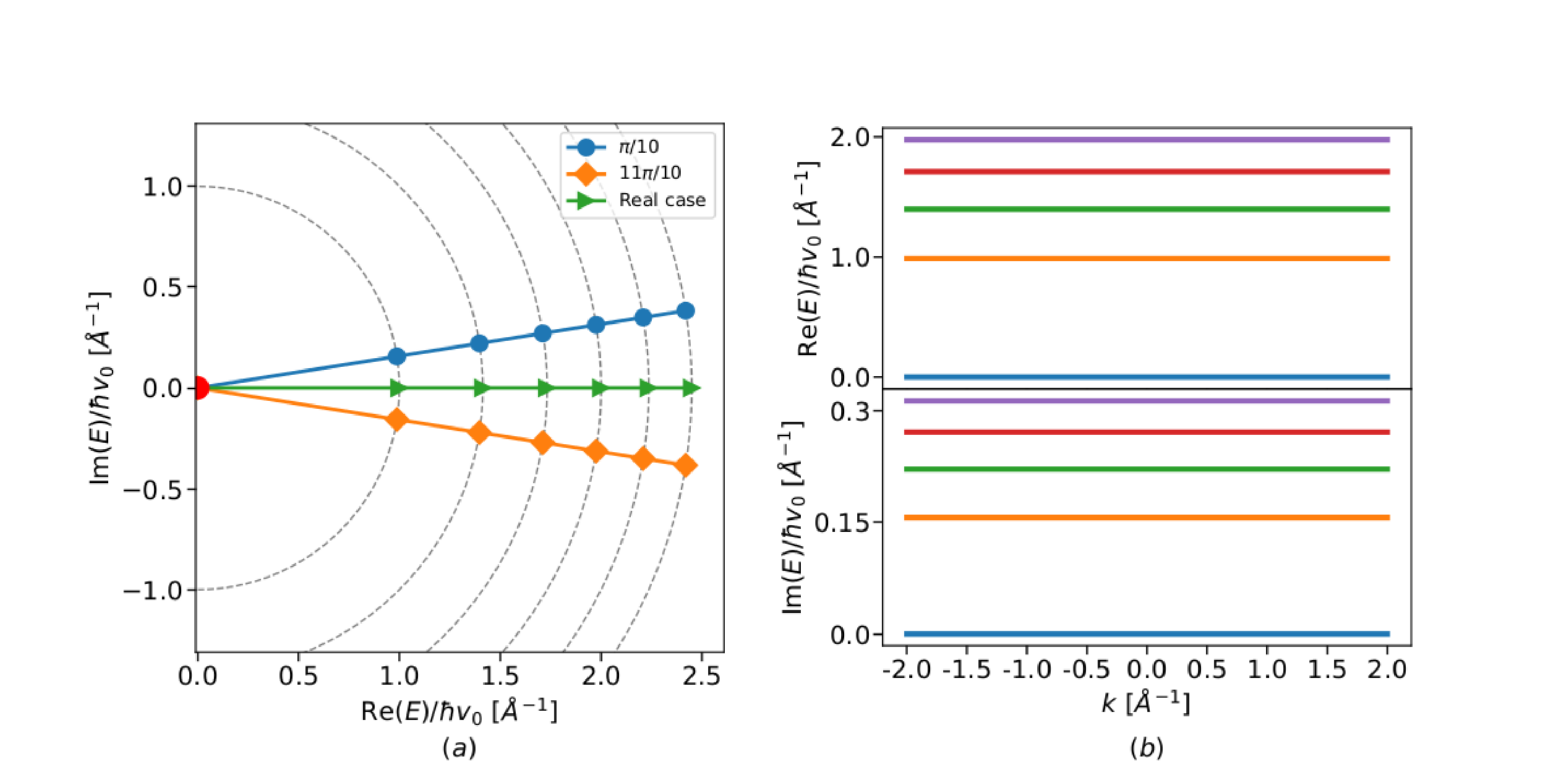}
\caption{(a) First energy eigenvalues in the complex plane for the constant magnetic profile with three different angles $\theta$. The ground state is the same for all these $\theta$-values and it is drawn as a red circle at the origin; the other potential parameters were taken as $|\omega| = k = 1$. (b) Real (top) and imaginary (bottom) part of the first five energy eigenvalues as functions of $k$ for $|\omega| = 1$ and $\theta = \pi/10$.} \label{F-2}
\end{center}
\end{figure}

\begin{figure}[t]
\begin{center}
\includegraphics[scale=0.45]{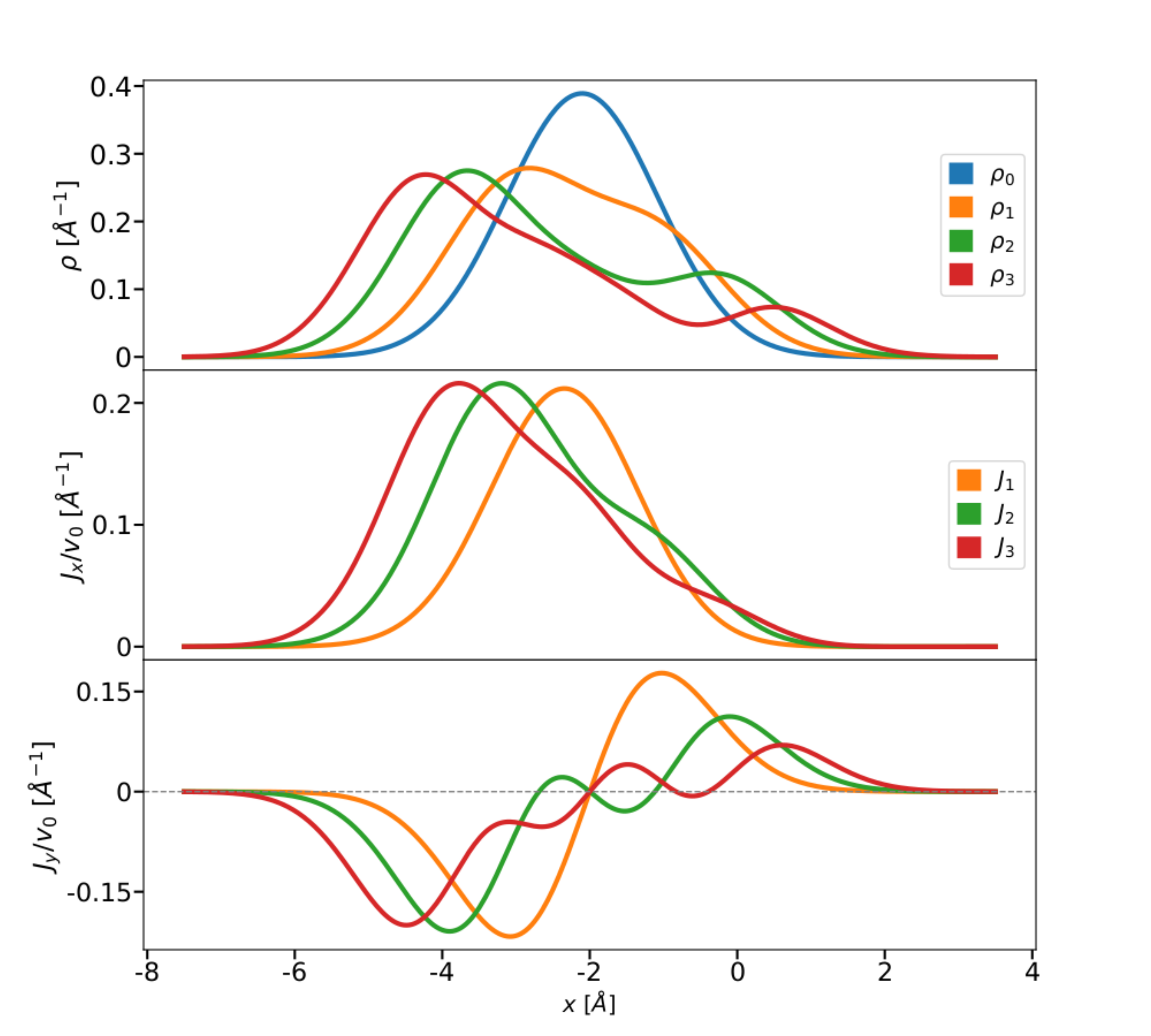}
\caption{Probability density (top), current density in the $x$-direction (middle) and in the $y$-direction (bottom) for the constant magnetic field. The potential parameters taken are $|\omega| = k = 1$ and $\theta = \pi/10$.} \label{F-3}
\end{center}
\end{figure}

The first magnetic profile we will consider is constant, i.e., $\mathbf{B}(x) = B \mathbf{e}_{z}$, $B\in \mathbb{C}$. In the Landau gauge the vector potential is $\mathbf{A}(x) = xB \mathbf{e}_{y}$. Substituting this expression in equation~\eqref{superpotential} we get $\text{w}(x) = k  + \omega\, x/2$ with $\omega = 2eB/c\hbar\in\mathbb{C}$, and the auxiliary SUSY partner potentials become
\begin{equation} \label{E-20}
\begin{aligned}
& V^{-}(x) =  \frac{\omega^{2}}{4}\left(x + \frac{2k}{\omega}\right)^{2} - \frac{\omega}{2},
\\
& V^{+}(x) =  \frac{\omega^{2}}{4}\left(x + \frac{2k}{\omega}\right)^{2} + \frac{\omega}{2}.
\end{aligned}
\end{equation}
These are called complex harmonic oscillators \cite{JuanCarlos2015}, whose real and imaginary parts can be obser{\-v}ed in figure~\ref{F-1}. The corresponding eigenfunctions are given by
\begin{equation} \label{E-21}
\psi^{\pm}_{n}(x) = 
\begin{cases}
c_{n}e^{-\frac{\zeta^{2}}{2}}\mathcal{H}_{n}\left[\zeta\right],\quad -\frac{\pi}{2} < \theta < \frac{\pi}{2},
\\
c_{n}e^{-\frac{\xi^{2}}{2}}\mathcal{H}_{n}\left[\xi\right],\quad \frac{\pi}{2} < \theta < \frac{3\pi}{2},
\end{cases}
\end{equation}
with $n$ being a non-negative integer, $\zeta = \sqrt{\omega/2}\left(x + 2k/\omega\right)$, $\xi = \sqrt{-\omega/2}\left(x - 2k/\omega\right)$, $\omega = |\omega|e^{i\theta}$ and $\mathcal{H}_{n}(\zeta)$ is a Hermite polynomial of degree $n$ and complex argument \cite{JuanCarlos2015}; we are denoting $\sqrt{\omega}=\sqrt{|\omega|}e^{i\theta/2}$ and $\sqrt{-\omega}=\sqrt{|\omega|}e^{i(\pi-\theta)/2}$. The eigenvalues for the potentials~\eqref{E-20} turn out to be 
\begin{equation} \label{E-22}
\varepsilon^{-}_{0} = 0,\quad \varepsilon^{-}_{n} = \varepsilon^{+}_{n-1} = \pm n\omega, 
\end{equation} 
where $n$ is a natural number, the upper sign $+$ is taken for $-\pi/2 < \theta < \pi/2$ and the lower sign $-$ for $\pi/2 < \theta < 3\pi/2$. Thus, the electron energies~\eqref{E-16} for graphene in a constant complex magnetic field can be written as follows:
\begin{equation} \label{E-23}
E_{n} = \hbar v_{0}\sqrt{\pm n\omega},
\end{equation}
whose norms coincide with the result for the real case deduced in \cite{Kuru2009} but now they are rotated in the complex plane by an angle $\theta/2$ with respect to the positive real line (see figure~\ref{F-2}(a)). In that plot it can be observed as well concentric circumferences of radius $R \propto \sqrt{n|\omega|}$ centered at the origin on which the energy $E_{n}$ is located regardless of the angle $\theta$. This leads us to conclude that despite its complex nature, for a fixed angle $\theta$ the spectrum of $H$ is ordered in the standard way. Moreover, $\text{Sp}\left(H\right)$ is infinite discrete, and its energies do not depend on $k$. In figure \ref{F-2}(b) it is shown the real and imaginary parts of the first energy eigenvalues as functions of $k$. The square-integrability of $\Psi_{n}(x,y)$ does not impose any constraint to the norm of $\omega$, but it does on its argument $\theta$, as it is shown in equation~\eqref{E-21}. Furthermore, when $\theta = \pm\pi/2$ the eigenfunctions $\psi^{\pm}_{n}(x)$ are not square-integrable, since in this case $V^\pm(x)$ in equation \eqref{E-20} become repulsive oscillator potentials \cite{Bermudez2013}, displaced by some imaginary quantities in the coordinate $x$ as well as in the energy origin, and thus the Hamiltonian $H$ does not have bound state solutions at all. The probability and current densities are drawn in figure~\ref{F-3} for the first four bound states. Note that the ``ground state'', for $n=0$, does not have associated any current density, since its upper entry is zero, as it is seen in equation~\eqref{E-16}.

\subsection{Trigonometric singular well}

\begin{figure}[t]
\begin{center}
\includegraphics[scale=0.75]{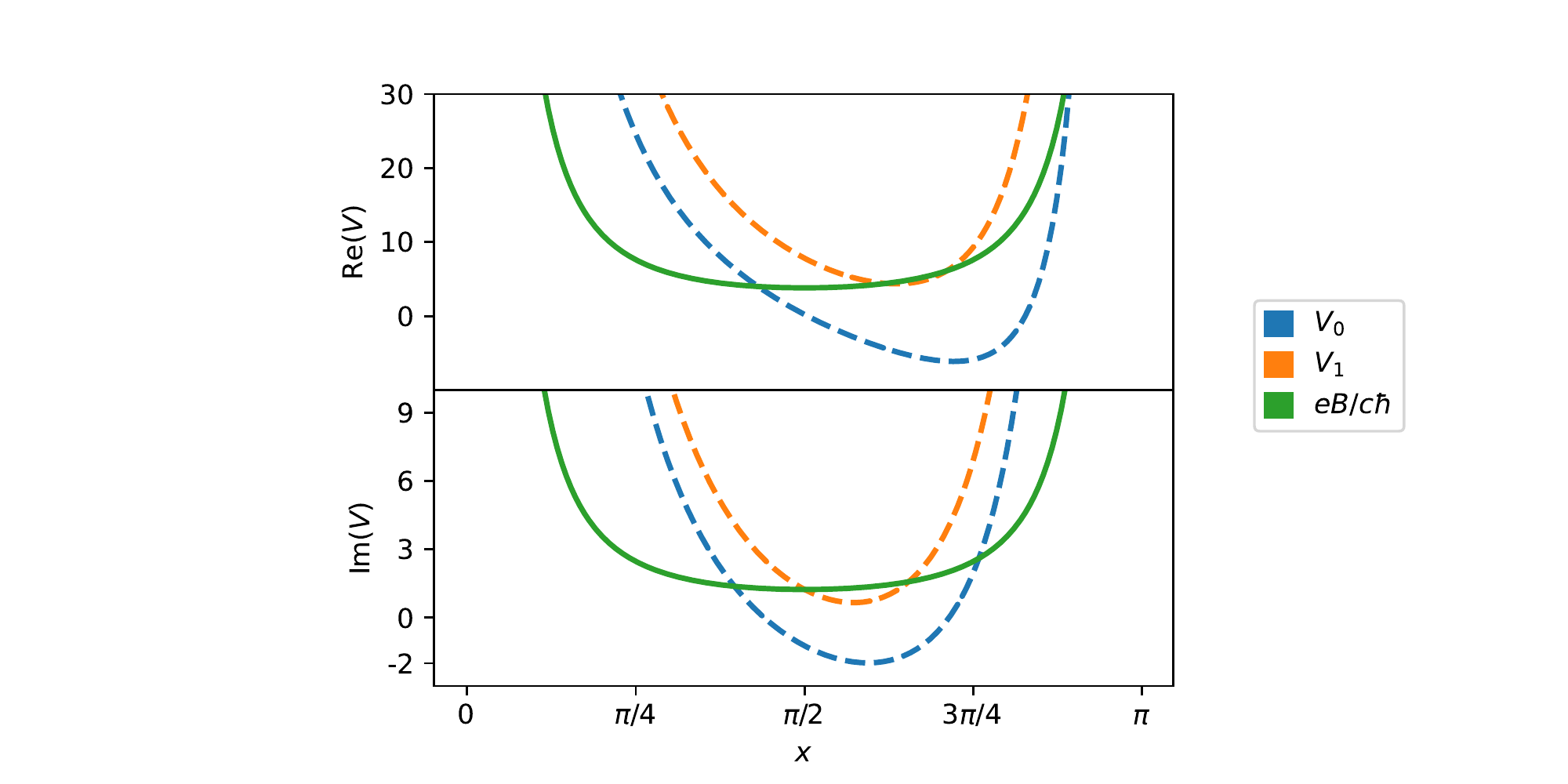}
\caption{Real (top) and imaginary part (bottom) of the complex trigonometric Rosen-Morse potentials and the corresponding magnetic field. The chosen potential parameters are $|D| = 4, \theta = \pi/10, k = -2, \mu = 1$.} \label{F-4}
\end{center}
\end{figure}

In this case a complex magnetic field of trigonometric form is taken, $\mathbf{B}(x) = B\csc^{2} \left(\mu x\right)\mathbf{e}_{z}$, $B\in \mathbb{C}$, $\mu\in\mathbb{R}^+$. The vector potential is given by $\mathbf{A}(x) = \left(-B/\mu\right) \cot (\mu x)\mathbf{e}_{y}$, thus it is straightforward to obtain the superpo{\-t}ential as $\text{w}(x) = k - D\cot (\mu x)$, with $D = eB/c\hbar\mu$. Hence, the auxiliary potentials now acquire the form 
\begin{equation} \label{E-24}
\begin{aligned}
& V^{-}(x) = D(D - \mu)\csc^{2} (\mu x) - 2Dk\cot (\mu x) + k^{2} - D^{2}, \\
& V^{+}(x) = D(D + \mu)\csc^{2} (\mu x) - 2Dk\cot (\mu x) + k^{2} - D^{2}. 
\end{aligned}
\end{equation} 
These expressions suggest to call the previous $V^\pm(x)$ as complex trigonometric Rosen-Morse potentials. Their real and imaginary parts are plotted in figure~\ref{F-4}. The corresponding eigenfunc{\-t}ions are given in terms of Jacobi polynomials $P_{n}^{(\alpha, \beta)}(\zeta)$ with complex argument and indexes, namely
\begin{equation} \label{E-25}
\psi^{j}_{n}(x) = c_{n} (-1)^{-\left(s_{j} + n\right)/2}(1 + \zeta^{2})^{-\left(s_{j} + n\right)/2}e^{r_{j}\text{arccot} (\zeta)}P_{n}^{\left(-s_{j} - n -ir_{j}, -s_{j} - n + ir_{j}\right)}(i\zeta),\quad j = \pm, 
\end{equation}
where $s_{-} = D/\mu$, $s_{+} = s_{-} + 1$, $r_{-} = -kD/\mu(D + n\mu)$, $r_{+} = -kD/\mu(D + \mu + n\mu)$, $\zeta = \cot (\mu x)$ and $n$ is a non-negative integer. Using now the polar form $D=|D|e^{i\theta}$, it turns out that the square-integrability of these eigenfunctions is limited to the right side of the complex plain, where $-\pi/2 < \theta < \pi/2$. The spectra of the Hamiltonians $H^{\pm}$ consist of the complex eigenvalues
\begin{equation} \label{E-26}
\varepsilon^{-}_{0} = 0,\quad \varepsilon^{-}_{n} = \varepsilon_{n-1}^{+} = k^{2} - D^{2} + (D + n\mu)^{2} - \frac{k^{2}D^{2}}{(D + n\mu)^{2}}, \quad n\in \mathbb{N}.
\end{equation}   
Substituting these expression in equation~\eqref{E-16} the electron energies for the complex trigono{\-m}etric singular magnetic field turn out to be 
\begin{equation} \label{E-27}
E_{n} = \hbar v_{0}\sqrt{k^{2} - D^{2} + (D + n\mu)^{2} - \frac{k^{2}D^{2}}{(D + n\mu)^{2}}}.
\end{equation} 

\begin{figure}[t]
\begin{center}
\includegraphics[scale=0.5]{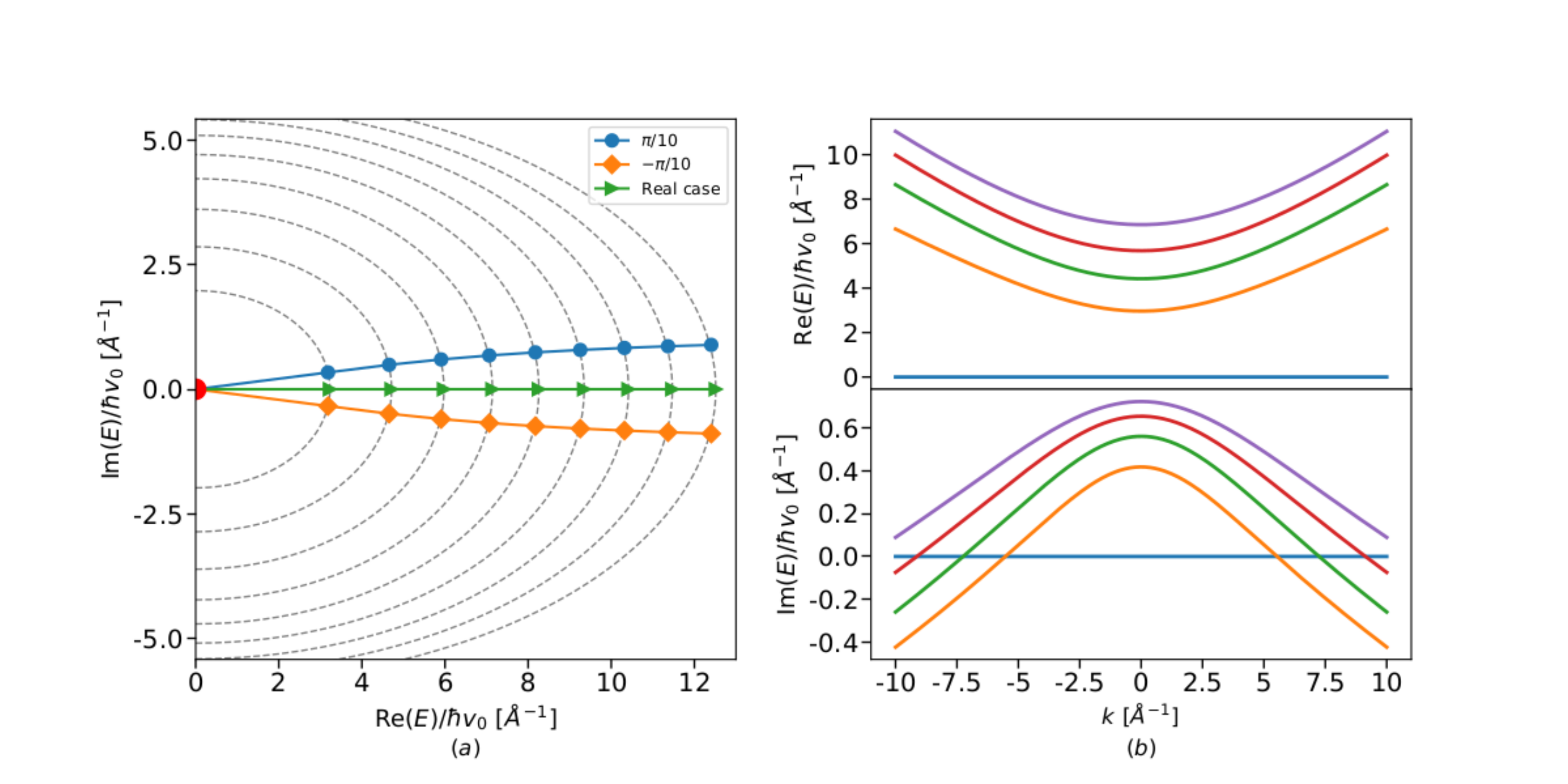}
\caption{(a) First electron energies $E_n$ for the trigonometric singular well in the complex plane with three different angles. The ground state is marked as a red circle at the origin, and the other potential parameters are $|D| = 4, k = -2, \mu = 1$. (b) Real (top) and imaginary part (bottom) of $E_n$ as functions of $k$ for $|D| = 4, \theta = \pi/10, \mu = 1$.} \label{F-5}
\end{center}
\end{figure}

\begin{figure}[t]
\begin{center}
\includegraphics[scale=0.6]{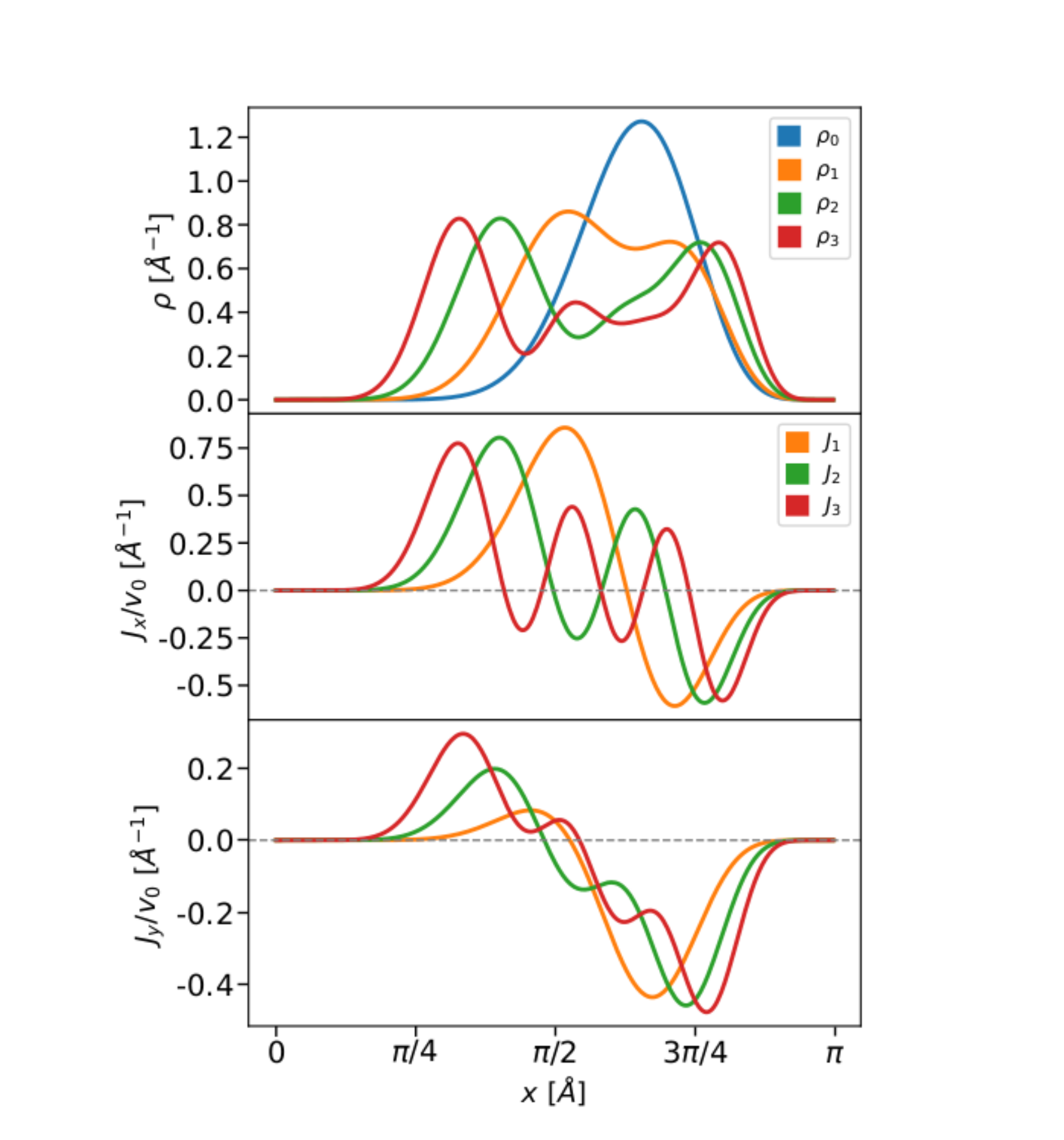}
\caption{Probability density (top), current density in $x$-direction (middle) and $y$-direction (bottom) in the case of a trigonometric singular well. The potential parameters taken are $|D| = 4, \theta = \pi/10, k = -2$ and $\mu = 1$.}  \label{F-6}
\end{center}
\end{figure}

We must mention that the norm of $E_{n}$ is different in general from the result for the real case shown in \cite{Kuru2009}, except in the case with $\theta = 0$. In fact, the argument of $E_n$ depends in a non-trivial way of $\theta$, and it shows also a strong dependence on the potential parameters. The first electron energies on the complex plane are shown in figure \ref{F-5}(a). It can be observed as well concentric ellipses centered at the origin, with the energy $E_{n}$ belonging to the ellipse whose semi-major axis coincides with the $n$-th energy in the real case regardless of the value of $\theta$. As in the previous case this fact implies that, for a fixed angle $\theta$, $\text{Sp}\left(H\right)$ is ordered in the standard way. However, this happens only in the interval $(-k_{0},k_{0})$, where $k_{0}$ is such that $\text{Im}(E_{1}(k_{0})) = 0$. Hence, the spectrum of $H$ turns out to be infinite discrete, as it can be seen in figure \ref{F-5}(b). Lastly, plots of the probability and current densities are displayed in figure~\ref{F-6} for the first four bound states. 

\subsection{Exponentially decaying magnetic field}

\begin{figure}[ht]
\begin{center}
\includegraphics[scale=0.75]{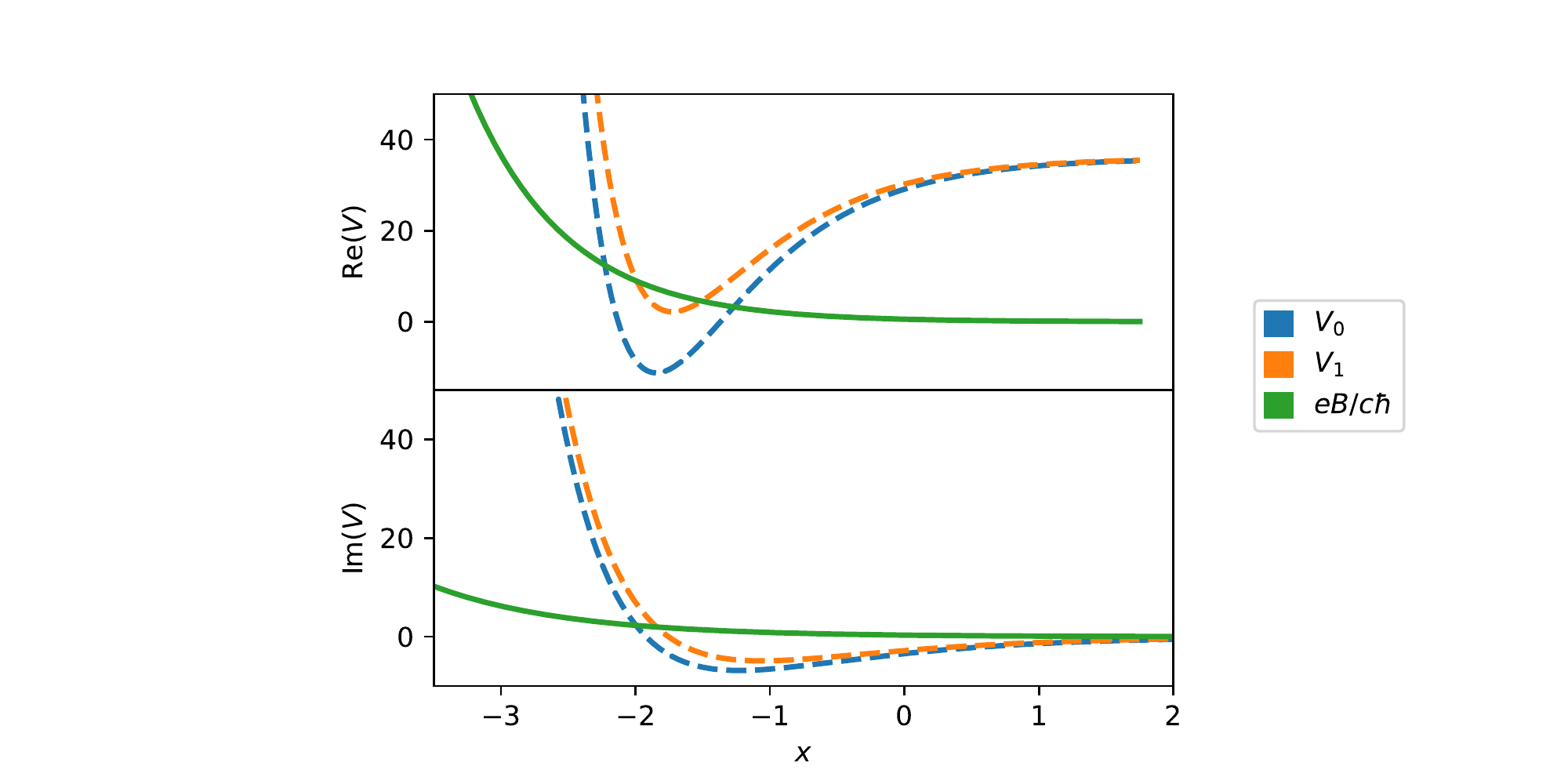}
\caption{Real (top) and imaginary part (bottom) of the complex Morse potentials and the corresponding magnetic field. The potential parameters taken are $|D| = 1, \theta = \pi/10, k = 6, \mu = 1$.} \label{F-7}
\end{center}
\end{figure}

Our last example is an exponentially decaying complex magnetic field $\mathbf{B}(x) =  Be^{-\mu x}\mathbf{e}_{z}$, $B\in \mathbb{C}$, $\mu\in\mathbb{R}^+$, whose vector potential is $\mathbf{A}(x) = -(B/\mu)e^{-\mu x} \mathbf{e}_{y}$. In agreement with equation~\eqref{superpotential}, the superpotential is given by $\text{w}(x) = k - De^{-\mu x}$, $D = eB/c\hbar\mu$. Inserting this expression in equation~\eqref{E-10}, we get the auxiliary SUSY partner potentials 
\begin{equation} \label{E-28}
\begin{aligned}
& V^{-} = k^{2} + D^{2}e^{-2\mu x} - 2D\left(k + \frac{\mu}{2}\right)e^{-\mu x}, \\
& V^{+} = k^{2} + D^{2}e^{-2\mu x} - 2D\left(k - \frac{\mu}{2}\right)e^{-\mu x},
\end{aligned}
\end{equation}
which are the Morse potentials but with the parameter $D$ being now complex. Their real and imaginary parts are shown in figure~\ref{F-7}. These potentials are as well exaclty solvable, the correspon{\-d}ing eigenfuntions are given by
\begin{equation} \label{E-29}
\psi^{j}_{n}(x) = c_{n}(\zeta)^{s_{j} - n}e^{-\frac{\zeta}{2}}L_{n}^{2(s_{j} - n)}(\zeta),\quad j = \pm,
\end{equation} 
where $s_{-} = k/\mu$, $s_{+} = s_{-} - 1$, $\zeta = (2D/\mu)e^{-\mu x}$, $n$ is a non-negative integer and $L_{n}^{\lambda}(\zeta)$ is an associated Laguerre polynomial of complex argument. The polar form $D = |D|e^{i\theta}$ allows us to deduce the square-integrability conditions: $-\pi/2 < \theta < \pi/2$ and $k > n\mu$. The corresponding eigenvalues are
\begin{equation} \label{E-30}
\varepsilon^{-}_{0} = 0,\quad \varepsilon_{n}^{-} = \varepsilon_{n - 1}^{+} = k^{2}-(k - n\mu)^{2},
\end{equation}
with $n$ being a natural number. It is worth stressing that the spectra of $H^\pm$ are real, since unlike the previous cases now these Hamiltonians are pseudo-hermitian \cite{Mostafazadeh2002}. Hence, the energy eigenvalues for the graphene electron in the exponentially decaying complex magnetic field turn out to be
\begin{equation} \label{E-31}
E_{n} = \hbar v_{0}\sqrt{k^{2}-(k - n\mu)^{2}}.
\end{equation}  

\begin{figure}[ht]
\begin{center}
\includegraphics[scale=0.75]{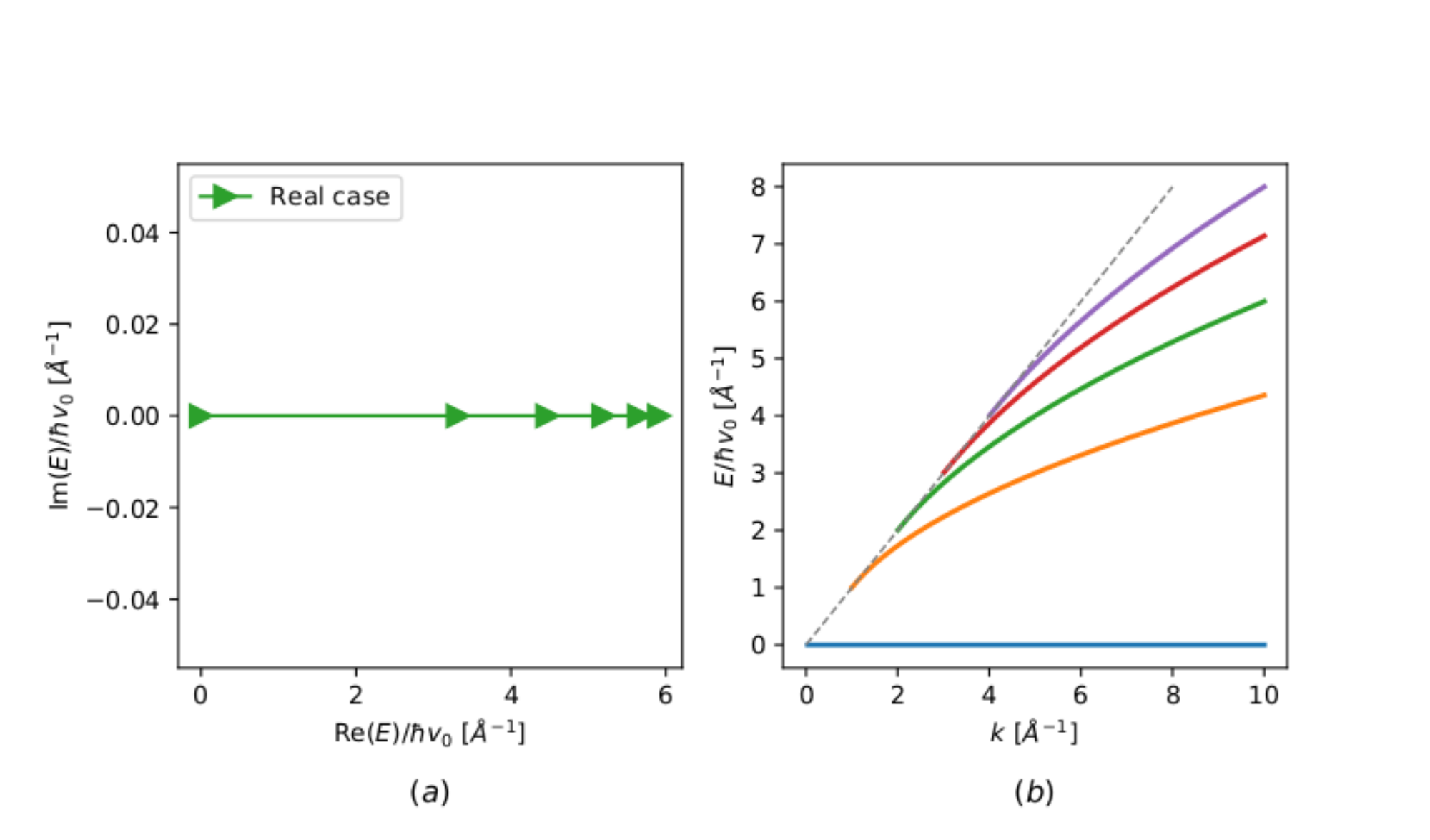}
\caption{(a) First electron energies $E_n$ for the exponentially decaying magnetic field in the complex plane for $k = 6, \mu = 1$. (b) Electron energies $E_n$ as functions of $k$ for the same $\mu$ value as in (a).} \label{F-8}
\end{center}
\end{figure}

\begin{figure}[ht]
\begin{center}
\includegraphics[scale=0.6]{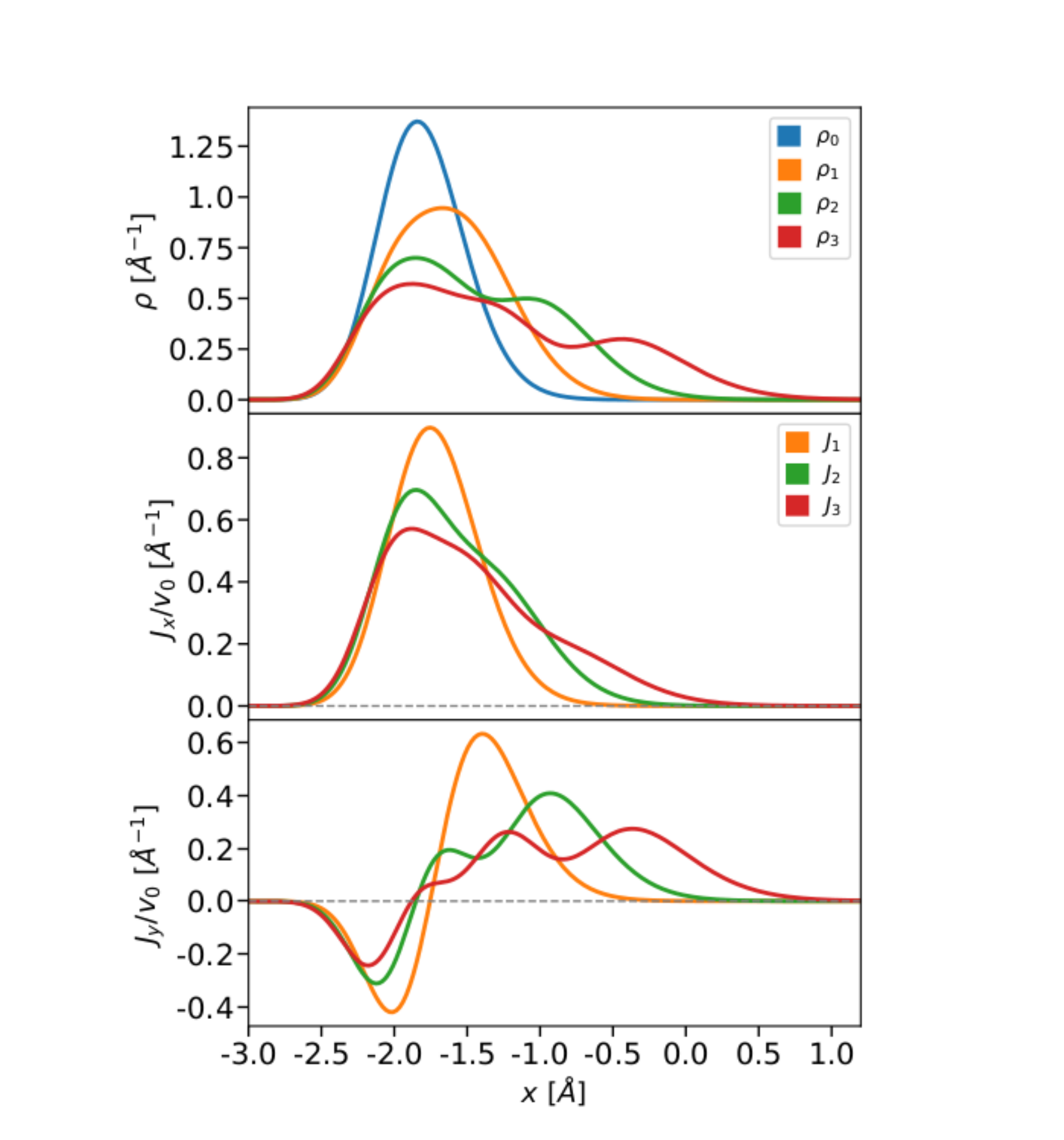}
\caption{Probability density (top), current density in the $x$-direction (middle) and the $y$-direction (bottom) for the exponentially decaying magnetic field. The potential parameters taken are $|D| = 1, \theta = \pi/10, k = 6$ and $\mu = 1$.}  \label{F-9}
\end{center}
\end{figure}

Note that, in this example, Sp($H$) coincides exactly with the spectrum for the real case address{\-e}d in \cite{Kuru2009}. Such spectrum is finite discrete, since once the parameters $k$ and $\mu$ are fixed, the condition $k > n\mu$ limits the number of square-integrable eigenfunctions and hence the number of allowed electron energies (see figure~\ref{F-8}). Moreover, there is an enveloping line, also shown in figure~\ref{F-8}(b), whose slope (equal to $v_{0}$) represents the average $y$-velocity. This line separates the $k$-domain into two subsets, one where there are bound states and the other where there are just scattering states. Finally, the probability and current densities are plotted in figure~\ref{F-9}.

\section{Discussion} \label{S4}  

It is interesting to observe that there are some $x$-points for which the non-trivial imaginary parts of $eB(x)/c\hbar$ and $V^+(x)$ are equal, as it is shown in figures~\ref{F-1}, \ref{F-4} and \ref{F-7}. If we denote as $\chi$ one of these points, it turns out that $\text{Im}[\text{w}^{2}(\chi)] = 0$ in order to fulfill equation~\eqref{E-10}, and this implies that $\text{Re}[\text{w}(\chi)] = 0$. Let us recall now a classical quantity, the {`\it kinematical'} momentum along $y$-direction given by $\Pi_{y} = p_{y} + (e/c)A$. Since the canonical momentum $p_{y}$ is a constant of motion, it follows that $\text{Re}[\Pi_{y}(\chi)] = \hbar\text{Re}[\text{w}(\chi)] = 0$. It is worth noticing that the maximum of the ground state probability density appears at the point $\chi$, and the latter depends on the angle $\theta$ (see figures~\ref{F-3},\ref{F-6} and \ref{F-9}).    

On the other hand, since the Hamiltonian \eqref{E-2} is non-hermitian, its eigenvalues are not necessari{\-l}y real. In fact, it can be written as $H/v_{0} = \sigma_{x}p_{x} + \sigma_{y}[p_{y} + (e/c)A(x)]$. Now, if we express the vector potential in polar form it turns out that 
\begin{equation} \label{E-32}
H = H_{R} + iv_{0}(e/c)\sigma_{y}|A(x)|\sin\theta,
\end{equation}
where $H_{R}= v_{0}\left[\sigma_{x}p_{x} + \sigma_{y}\left(p_{y} + (e/c)|A(x)|\cos\theta\right)\right]$ is a hermitian operator whose eigenvalues are real, which is similar to the Hamiltonian addressed in \cite{Kuru2009}. The second term of equation~\eqref{E-32} is an anti-hermitian operator whose eigenvalues are purely imaginary. In order to understand the nature of this term, let us remember that the Dirac-Weyl equation in graphene describes a massless {\it pseudo-spin} $1/2$ particle, where {\it pseudo-spin `up'} means that the electron is in the sublattice B and {\it `down'} in the sublattice A. In terms of the pseudo-spin ladder operators $S_{\pm} = S_{x} \pm iS_{y}$ it follows that the second term can be written as $(ev_{0}/c\hbar)\left(S_{+} - S_{-}\right)|A(x)|\sin\theta$. It induces a pseudo-spin rotation, and it is analogous to the corresponding term that appears in the Hamiltonian describing non-uniformly strained graphene (see \cite{Maurice2015}). By sticking to this analogy, in \cite{Vozmediano2013} this anti-hermitian term is associated to the layer curvature induced by strain, while in the case analyzed here the analogous term is proportional to the imaginary part of the vector potential, however the issue is to find a phenomenon that could be associated with it. Then, since the Hamiltonian \eqref{E-32} is time-independent, the time evolution of the total probability associated to any eigenstate has an exponential factor which depends on the imaginary part of its eigenvalue $E_{n}$, namely,
\begin{equation} \label{E-33}
\mathscr{P}_{T}(t) = \langle\Psi_{n}(t)|\Psi_{n}(t)\rangle = e^{2\frac{\text{Im}[E_{n}]}{\hbar}t}\langle\Psi_{n}(0)|\Psi_{n}(0)\rangle.
\end{equation}
A small probability increase (decrease) occurs when the exponent $2(\text{Im}[E_{n}]/\hbar)t\ll 1$, which happens for approximate times inversely proportional to the imaginary part of the eigenva{\-l}ue. Using the polar form $E_{n} = |E_{n}|e^{i\phi_n}$ for the first bound states, it turns out that for $\phi_n\ll 1$ we obtain long enough times in order to guarantee the probability conservation except for a small perturbation term. Thus, the anti-hermitian term in the Hamiltonian \eqref{E-32} can be seen as a perturbation describing the loss or gain of charger carriers in the graphene sublattices. Therefore, one could consider the graphene in a real magnetic field orthogonal to its surface, trying to model a non-conservative system due to the interaction of the {\it pseudo-spin} electron with the magnetic field by means of the exactly solvable non-hermitian Hamiltonian~\eqref{E-2}, and taking the expansion of the eigenvalues $E_{n}$ in powers of its argument $\phi_n$ one could find the corresponding energy corrections. Furthermore, the probability and current densities will be now modified as compared with the real case. It is worth noticing that despite the Hamiltonian is non-hermitian, this property does not ensure that its eigenvalues are complex, as it was seen in the third example worked out in the previous section where the eigenvalues were real, and thus the total probability of its eigenstates was conserved.

Finally, we must mention an interesting case where the anti-hermitian term in the Hamiltonian \eqref{E-32} is not a perturbation around $\theta=0$: in the limit $\theta\rightarrow\pi/2$ our Hamiltonian describes the free graphene \cite{Ferreira2011} plus a term of interaction between the {\it pseudo-spin} electron and a purely imaginary magnetic field. All the magnetic profiles used in section \ref{S3} lead us to auxiliary potentials without bound states in this limit. However, if we choose an argument $\theta\approx\pi/2$ we will get potentials with ``weak'' bound states, whose probability densities have a pronounced maximum that diverges in the limit when $\theta$ tends to $\pi/2$.  

\section{Conclusions} \label{S5}

Exact analytic solutions for non-hermitian Hamiltonian $H$ describing an electron in graphene interac{\-t}ing with external complex magnetic fields were found. It is worth noticing that $H$ can be expressed as the addition of a hermitian operator plus a non-hermitian term, the last causing pseudo-spin rotations. Although, we have solved the eigenvalue problem for the Hamiltonian \eqref{E-2} assuming that a complex magnetic field in principle can be produced, a possible physical interpretation could be that a real magnetic field, with the same amplitude as the complex one, is being applied to graphene, and the angle $\theta$ is a parameter allowing us to introduce in the Hamiltonian a term describing the loss or gain of charge carriers induced by the interaction between the {\it pseudo-spin} electron and the magnetic field. Due to the complex nature of the problem, there are important differences in some physical quantities with respect to the real case, as the probabili{\-t}y and current densities. In particular, the current density along $x$-direction turns out to be non-null, in contrast to the result found in \cite{Kuru2009} for the real case. Furthermore, now the ground state probability density acquires its maximum at the point where the imaginary parts of $eB/\hbar c$ and $V^{+}$ are equal.

\section*{Acknowledgements}
This work was supported by CONACYT (Mexico), project FORDECYT-PRONACES/61533/20\\20. JDGM especially thanks to Conacyt for the economic support through the PhD scholarship.

\bibliography{complex_fields}
\bibliographystyle{unsrt}
\addcontentsline{toc}{section}{References} 
\end{document}